\input ./style/arxiv-general.cfg
\documentclass[aoas,MSNbibl,nameyear,seceqn,dvips]{arximspdf}
\makeatletter
   \@ifpackageloaded{graphicx}{}{\usepackage{graphicx}}
\makeatother
\usepackage{multirow}

\doi{10.1214/15-AOAS860}
\volume{9}
\issue{4}
\pubyear{2015}
\firstpage{2215}
\lastpage{2236}
\docsubty{FLA}

\begin{document}
\begin{frontmatter}

\title{On the analysis of tuberculosis studies with intermittent
missing sputum data\thanksref{T1}}
\runtitle{Tuberculosis studies with missing sputum data}

\begin{aug}
\author[A]{\fnms{Daniel}~\snm{Scharfstein}\corref{}\thanksref{M1}\ead[label=e1]{dscharf@jhu.edu}},
\author[B]{\fnms{Andrea}~\snm{Rotnitzky}\thanksref{M2}\ead[label=e2]{arotnitzky@utdt.edu}},
\author[C]{\fnms{Maria}~\snm{Abraham}\thanksref{M3}\ead[label=e3]{maria.abraham4@gmail.com}},
\author[A]{\fnms{Aidan}~\snm{McDermott}\thanksref{M1}\ead[label=e4]{amcderm1@jhu.edu}},
\author[D]{\fnms{Richard}~\snm{Chaisson}\thanksref{M1}\ead[label=e5]{rchaiss@jhmi.edu}}
\and
\author[E]{\fnms{Lawrence}~\snm{Geiter}\thanksref{M4}\ead[label=e6]{Lawrence.Geiter@otsuka-us.com}}
\runauthor{D. Scharfstein et al.}
\affiliation{Johns Hopkins University\thanksmark{M1},
Universidad Torcuato Di Tella\thanksmark{M2},
Statistics Collaborative\thanksmark{M3} and
Otsuka Novel Products\thanksmark{M4}}
\address[A]{D. Scharfstein\\
A. McDermott\\
Department of Biostatistics\\
Johns Hopkins Bloomberg School\\
\quad of Public Health\\
615 North Wolfe Street\\
Baltimore, Maryland 21205\\
USA\\
\printead{e1}\\
\phantom{E-mail: }\printead*{e4}}
\address[B]{A. Rotnitzky\\
CONICET\\
Department of Economics\\
Universidad Torcuato Di Tella\\
Saenz Valiente 1010\\
1428 Buenos Aires\\
Argentina\\
\printead{e2}}
\address[C]{M. Abraham\\
Statistics Collaborative\\
1625 Massachusetts Ave NW\\
Suite 600\\
Washington, DC 20036\\
USA\\
\printead{e3}}
\address[D]{R. Chaisson\\
Johns Hopkins Center\\
\quad for Tuberculosis Research\hspace*{8pt}\\
1550 Orleans St., 1M.08\\
Baltimore, Maryland 21231\\
USA\\
\printead{e5}}
\address[E]{L. Geiter\\
Otsuka Novel Products-TB\\
Otsuka Pharmaceutical Development\\
\quad and Commercialization, Inc.\\
2440 Research Boulevard\\
Rockville, Maryland 20850\\
USA\\
\printead{e6}}
\end{aug}
\thankstext{T1}{Supported in part by NIH Grants CA85295, CA183854,
AI051164, P30 MH066247
and Otsuka Pharmaceutical
Development and Commercialization, Inc. and contracts from the Food and
Drug Administration and the Patient-Centered Outcomes Research Institute.}

%
\received{\smonth{5} \syear{2013}}
%
\revised{\smonth{7} \syear{2015}}

%
\begin{abstract}
In randomized studies evaluating treatments for tuberculosis (TB),
individuals are scheduled to be routinely evaluated for the
presence of TB using sputum cultures. One important endpoint in such
studies is the time of culture conversion, the first visit~at which a patient's sputum culture is negative and remains negative.
This article addresses how
to draw inference about treatment effects when sputum cultures are
intermittently missing on some patients. We discuss inference
under a novel benchmark assumption and under a class of assumptions
indexed by a treatment-specific sensitivity parameter that quantify departures
from the benchmark assumption. We motivate and illustrate our approach
using data from a randomized trial comparing the effectiveness of two treatments
for adult TB patients in Brazil.
\end{abstract}

%
\end{frontmatter}

\setcounter{footnote}{1}
\section{Introduction}\label{sec1}

In the design of randomized studies evaluating competing treatments for
patients with tuberculosis (TB), it is common to culture sputum for the
presence of TB at regularly scheduled clinic visits over a specified time
horizon. A primary goal in such studies is to estimate the treatment-specific
distribution of the time of culture conversion [\citet{emea2010}].
Culture conversion is said to have occurred for a patient at a given
visit~if the sputum cultures for that visit~and
all subsequent visits are negative. A key complication in the analysis
arises when culture results are missing at some
visits, because the culture was contaminated, the patient
could not produce sputum, or the patient did not show up. Culture
conversion status at a given visit~is unknown
when from that visit~onward at least one culture result is missing
\textit{and} all recorded culture results are negative. For a given
patient, the set of visits with unknown
culture conversion status is empty, or it consists of either a single
visit~or a set of consecutive visits.
If the set is not empty, the time of culture conversion will be known
to lie in an interval. The time may not be interval-censored in the
classical sense, however, because certain data configurations may
imply that culture conversion cannot occur at certain visit~times
within the interval. To distinguish this data structure from
classic interval censoring, we refer to the set of feasible times that
are compatible with an individual's data
as the coarsening set. The coarsening set can include a single point
or a set of points.

The treatment-specific distribution of time
of culture conversion is not identified without untestable
assumptions about the distribution of culture conversion status within
the coarsening sets. There are countless ways of imposing such assumptions.
The ``worst-case'' and ``best-case'' assumptions, leading to bounds
on the treatment-specific distribution of time of culture conversion,
are that the missing culture associated with the latest visit~time at which
culture conversion status is unknown is positive and that the
missing cultures associated with all visit~times at which culture
conversion status is
unknown are negative, respectively. In considering alternative assumptions,
it is natural to condition on as much of the
relevant data as possible. In addition to
conditioning on observed culture results, it is natural to condition on
auxiliary
factors that are associated with the unknown
results inside the coarsening set.

Most TB studies collect a key auxiliary time-varying variable. Sputum
specimens are also evaluated by smear, and the results of the smear may
be available when a culture is contaminated. Sputum smear is a less
reliable assessment of clinical tuberculosis than sputum culture. It
relies on the visualization of the bacteria through a microscope after
staining of sputum with dyes that allows the microscopist to see
so-called acid-fast TB bacteria. Sensitivity of the sputum smear is
about 0.5, but it can vary by staining method used and clinical
population. 
In contrast, sputum culture involves breaking down the sputum (which
is very viscous), decontaminating the specimen to kill bacteria other
than the mycobacteria, and inoculating it on culture media where the
bacilli can grow. Sensitivity of the sputum culture is about 0.8--0.85
[\citet{american2000diagnostic}]. Roughly 65\% of those with a positive
culture are expected to have a positive smear, and nearly 100\% of
those with a positive smear are expected to have a positive culture
[\citet{american2000diagnostic}].

Another important auxiliary variable is baseline cavitation status.
Patients with pulmonary tuberculosis who have cavities seen on a chest
radiograph (cavitary tuberculosis) are more likely to have positive
sputum smears, as they harbor larger numbers of tubercle bacilli than
patients without cavities. It generally takes longer for patients with
cavitary tuberculosis to convert their smears and cultures to negative
during treatment, as they have a larger bacterial load and the therapy
must kill more organisms.


This\vspace*{1pt} article is motivated by data from a randomized TB study previously
analyzed by \citet{conde2009}.\hskip.2pt\footnote{The data set provided to
us differs slightly from that of \citet{conde2009}. There are
small differences in the number of observed cultures and the number of
observed negative cultures at each week. All analyses reported in this
article are based on the data provided to us.} This phase II, double-blind,
randomized trial compared moxifloxacin vs. ethambutol in adults with
smear-positive tuberculosis at baseline in a hospital in Rio de Janeiro,
Brazil. All 170 patients randomized (85~to each treatment arm) into
the study were treated with a background regimen of isoniazid, rifampicin
and pyrazinamide. Patients with a negative or contaminated smear or
with drug-resistant \textit{Mycobacterium tuberculosis}
at baseline were excluded from the
analysis, resulting in an analysis sample of 74 and 72 patients in the
moxifloxacin and ethambutol groups,
respectively. Treatment was scheduled to be given five days per week
and was to
be directly observed by study personnel. Sputum specimens
(spontaneous or induced) were scheduled to be collected at baseline
and every week for 8 weeks. The specimens were to be evaluated by both
smear and culture testing.
In this study, 55.4\% and 62.5\% of patients in the moxifloxacin and
ethambutol arms, respectively, had complete culture data through week 8.
Time of culture conversion could be determined for 64.9\% and 72.2\% of
patients in these arms; the remaining patients had their time of culture
conversion coarsened.

In this article, we develop a method that estimates the
treatment-specific distribution of time of culture conversion under a class
of assumptions on the distribution of the coarsened time of culture conversion
that conditions on \emph{all of the relevant available data}, including sputum
cultures, sputum smears and baseline data. Each assumption in the class is
indexed by a treatment-specific sensitivity-analysis parameter which
quantifies the magnitude
of discrepancy from a specific benchmark assumption. In Section~\ref{sec2} we
provide a preview of our proposed method. In Section~\ref{sec3} we
discuss our modeling assumptions and approach to inference. Section~\ref{sec4}
presents an analysis
of data from the \citet{conde2009} study. The article concludes with
a discussion.

\section{Preview}\label{sec2}

Our approach starts by imposing nontestable assumptions that identify
the conditional distribution
of time to culture conversion given the data, for every data
configuration for which
time of culture conversion is unknown. The
crucial methodological challenge then is to make sensible
identifying assumptions. Because time of culture conversion is
determined by
the results of sputum cultures, these assumptions ultimately
identify the visit-specific probabilities of
the last positive sputum culture given the data. Lack of
culture conversion at a given visit~can be determined without full
knowledge of all subsequent results if the culture at that visit~is positive or at least one subsequent culture is observed to be
positive. It turns out that
with ``proper bookkeeping,'' we can achieve identifiability by imposing
assumptions that suffice to identify the distribution of time of
culture conversion
but do not fully identify the joint distribution of results across
visits. These conditions identify the reverse-time conditional hazards
of time of culture conversion. This section illustrates these issues by means
of an example.

%
\begin{table}[b]
\tabcolsep=0pt
\caption{Examples of patient data.
$-$ denotes negative culture,
$+$ denotes positive culture,
$U$ denotes unknown,
$I$ denotes a missing culture result that is irrelevant for determining culture conversion, and
$R$ denotes a missing culture result that is relevant for determining culture conversion}\label{example1}
\begin{tabular*}{\tablewidth}{@{\extracolsep{\fill}}@{}lcccccccccc@{}}
\hline
& & \multicolumn{8}{c}{\textbf{Visit}} & \multirow{3}{44pt}{\centering{\textbf{Coarsening set}}} \\[-6pt]
& & \multicolumn{8}{c}{\hrulefill}\\
\textbf{Line} & & \textbf{1} & \textbf{2} & \textbf{3} & \textbf{4} & \textbf{5} & \textbf{6} & \textbf{7} & \textbf{8} \\
\hline
\phantom{0}1 & Culture results & {\em I} & $+$ & {\em R} & $-$ & {\em R} & $-$ & $-$ & $-$ & \\
\phantom{0}2 & Conversion? & N & N & {\em U} & {\em U} & {\em U} & Y & Y & Y & $\{3,4,6 \}$ \\[3pt]
\phantom{0}3 & Culture results & {\em I} & $+$ & {\em I} & $-$ & $+$ & $-$ & $-$ & $-$ & \\
\phantom{0}4 & Conversion? & N & N & N & N & N & Y & Y & Y & $\{ 6 \}$\\[3pt]
\phantom{0}5 & Culture results & {\em I} & $+$ &{\em R} & $-$ & $-$ & $-$ & $-$ & $-$ & \\
\phantom{0}6 & Conversion? & N & N & U & Y & Y & Y & Y & Y & $\{ 3,4\}$\\[3pt]
\phantom{0}7 & Culture results & {\em I} & $+$ & $+$ & $-$ & $-$ & $-$ & $-$ & $-$ & \\
\phantom{0}8 & Conversion? & N & N & N & Y & Y & Y & Y & Y & $\{ 4 \}$ \\[3pt]
\phantom{0}9 & Culture results & {\em I} & $+$ & $-$ & $-$ & $-$ & $-$ & $-$ & $-$ & \\
10 & Conversion? & N & N & Y & Y & Y & Y & Y & Y & $\{ 3 \}$\\
\hline
\end{tabular*}
\end{table}

The sputum culture data collected on one patient in the
study illustrate the coarsening structure of culture
conversion status and the time of culture conversion. The sputum
culture data for this
patient, whom we call Mary, are displayed in the first
line of Table~\ref{example1}; the associated culture conversion
statuses are displayed
in line 2. Missing values are indicated by either {\em R} or {\em I},
depending on whether they are relevant or irrelevant for establishing
culture conversion status.
Mary has negative cultures
at visits 4, 6, 7 and 8 and a positive culture at visit~2. She is a
culture converter at visits 6, 7 and 8 (labeled ``Y'' in the second line),
known not to be a culture converter at visits 1 and 2 (labeled ``N'' in
the second
line) and has unknown culture conversion status at visits 3, 4 and 5
(labeled {\em U} in the second line). First, even though her culture
status at visit~1 (labeled {\em I} in the first line)
is missing, it is irrelevant for determining
whether she is a culture converter at that visit, because she has a
positive culture at visit~2 and therefore cannot be a
culture converter at visit~1. Second, missingness of cultures at visits
3 and 5 (labeled {\em R} in the first line) affects our ability to
determine her culture
conversion status at visits 3, 4 and 5 because all cultures after visit~5 are
negative and the culture at visit~4 is also
negative. So even though Mary has a negative culture at visit~4, her culture
conversion status is not known at that visit.
Third, the visits at which culture
conversion status is unknown are consecutive: the earliest and latest visits
are 3 and 5, respectively. Finally,
the coarsening set for time of culture conversion comprises visits 3, 4
and~6. This follows because
if the culture at visit~5 were positive, then the time of culture
conversion would be
visit~6, and if the culture at visit~5 were negative, the time of
culture conversion would be either
visit~3 or visit~4, depending on the result at visit~3.

To illustrate our approach for identifying the
conditional distribution of time of culture conversion given the data,
consider subset $A$, the subset of patients with the same observed data
as Mary.
Specifically, we must model the probability
that the cultures at visits 3 and 5 are both negative (in which case
the time of culture conversion is visit~3),
the probability that the culture at visit~3 is positive and the culture
at visit~5 is negative (in which
case the time of culture conversion is visit~4), and the probability
that the culture at visit~5 is positive
(in which case the time of culture conversion is visit~6). In modeling
these probabilities, it is natural to consider
two chronological factorizations. In forward time, we would need to
model the probability of a negative culture at visit~3,
the conditional probability of a negative culture at visit~5 given a
negative culture at visit~3, and the conditional probability of a
negative culture at
visit~5 given a positive culture at visit~3. In reverse time, we would
need to model (i) the probability of a positive culture
at visit~5 and (ii) the conditional probability of a positive culture
at visit~3 given a negative culture at visit~5. We use this
latter factorization as it requires fewer modeling assumptions.

We first turn to the task of imposing assumptions that identify (i).
Our identifying assumption specifies that (i) is the same as the probability
of a positive culture at visit~5 among patients with the same
data as in subset $A$, with the exception that they have an observed
culture at visit~5. The observed culture results for these patients
are depicted in lines 3 and 5. These patients have observed culture
conversion status as depicted in lines 4 and 6. The probability (i) is then
assumed to be equal to the ratio of the proportion of patients with
observed results as in line 3 to the sum of the proportions of
patients with observed results as in lines 3 and 5.

Next, we turn to the task of imposing assumptions that identify (ii). Our
identifying assumption specifies that (ii) is the same as the
probability of
a positive result at visit~3 among patients with the same pattern of
observed cultures as that in line~5, with the exception that they have an
observed culture at visit~3. The results of these patients are
depicted in lines 7 and 9 with associated culture conversion status in
lines 8 and 10. The probability (ii) is then assumed to be equal to the
ratio of the proportion of patients with observed cultures as in line
7 to the sum of the proportions of patients with observed cultures
as in lines 7 and 9.

In data sets of typical size, we will not be able to obtain reliable estimates
of the
proportion of patients who have specific patterns of observed data
because of
the curse of dimensionality. As a result, our inference will require
dimension-reduction assumptions. We will use fully
parametric models for the treatment-specific distributions of the
data. These models are described in Section~\ref{sec3.4}.

In Section~\ref{sec3.3} we evaluate the sensitivity of our results to our identifying
assumptions by conducting inference under a class of exponential tilt
deviations from the assumed conditional probabilities for the unobserved
culture conversion status.

\section{Formalization of the problem}\label{sec3}

Since we focus on inference about the time of culture conversion, separately
for each treatment arm, we consider, until Section~\ref{sec3.5}, only data from
one arm and suppress
notational dependence on treatment assignment.


\subsection{Data structure and notation}\label{sec3.1}

Let $X$ denote baseline cavitation status (1~for cavitation, 0
otherwise), $%
C_{k}$ denote the indicator that the culture is negative at visit~$k$
(1 for
negative, 0 for positive) and $S_{k}$ denote the indicator that the
smear is
negative at visit~$k$ (1 for negative, 0 for positive). Let $M_{k}^{c}$
and $%
M_{k}^{s}$ be the indicators that $C_{k}$ and $S_{k}$ are missing,
respectively (1 for missing, 0 for observed). The data recorded on an
individual at
visit~$k$ are a realization of the random vector $O_k =
(M_{k}^{c},C_{k}^{\mathrm{obs}},M_{k}^{s},S_{k}^{\mathrm{obs}})$, where $%
C_{k}^{\mathrm{obs}}=C_{k}$ if $C_{k}$ is observed and $C_{k}^{\mathrm{obs}}$ is empty
otherwise and $S_{k}^{\mathrm{obs}}$ is defined likewise.

Let $K$ denote the number of scheduled post-baseline visits. For any
collection of random vectors $\{W_{k}\}_{1\leq k\leq K}$, we
use the notation $\overline{W}_k = (W_1,\ldots, W_k)$.
With this notation, the data recorded on an individual throughout the entire
study are a realization of the random vector $\mathbf{O}=(X,\overline
{O}_K)$. It
is useful to denote the auxiliary data by $V
\equiv ( X, \overline{M}_{K}^{s},\overline{S}_{K}^{\mathrm{obs}} )$.
For a random variable or vector which is a function of $\mathbf{O}$
(e.g., $V$, $L$ and $R$), we
use lowercase notation (e.g., $v$, $l$ and $r$) to denote the
realization associated with a given realization $\mathbf{o}$ of $\mathbf{O}$.

Define time of culture conversion $T$ to be the earliest visit~such that
sputum cultures are negative from that visit~onward if such a visit~exists
and $T=K+1$ otherwise. With observed data $\mathbf{O}$ on a patient,
$T$ belongs either to a set
with a single visit~time (in which case $T$ is determined from $\mathbf{O}$)
or to a set with multiple visit~times, not necessarily consecutive.
We denote the coarsening set where $T$ is known to lie by ${\mathcal T}$.
Given $\mathbf{O}$, $T$ is determined (i.e., the set ${\mathcal T}$ has
one element) unless either:
\begin{longlist}[(ii)]
\item[(i)] the culture at visit~$K$ is missing, that is, $M_{K}^{c}=1$,
or
\item[(ii)] there exists a visit~$k<K$ with a missing culture, that is,
$M_{k}^{c}=1$, such that all subsequent visits have sputum cultures
that are
either negative or missing, that is, $M_{j}^{c}=1$ or $C_{j}^{\mathrm{obs}}=1$
for $j>k$.
\end{longlist}

When either (i) or (ii) occurs, ${\mathcal T}$ will have multiple visit~times. We denote the lowest visit~number in the set by $L$, which is
the earliest visit~$k$ where
the sputum cultures are either missing or negative at and subsequent to
visit~$k$.
We denote the largest visit~number in the set by $R+1$, where $R=K$ if
the culture at visit~$K$ is missing (i.e., $M_{K}^{c}=1$) and
$R = k$ $(k<K)$ if at visit~$k$ the sputum culture is missing and at all
subsequent visits the sputum cultures are recorded and
negative. The other times in the coarsening set include all visit~numbers $k$ such that
$L < k < R+1$ and $M_{k-1}^c = 1$. 


Formally, our inferential goal is to estimate the distribution of time of
culture conversion, that is, $P[T=k]$ for $k=1,\ldots,K$, based on $n$ i.i.d.
realizations of the vector $\mathbf{O}$. We use the subscript $i$ to denote
data for the $i$th individual.

In Section~\ref{sec3.2} we formally describe the identifying assumptions on
which our benchmark analysis relies. These assumptions were illustrated in
Section~\ref{sec2}. The assumptions are ``identifying'' in the sense that once
they are
imposed, we are able to express $P[ T=k ] $ for $k=1,\ldots,K $as a function
of the distribution of the observed data $\mathbf{O}$ and,
consequently, we can hope,
to estimate $P[ T=k] $ consistently. Subsequently, we
propose models for departures from the benchmark assumptions that form the
basis of our proposed sensitivity analysis. Specifically, our sensitivity
analysis consists of repeating estimation of $P[ T=k] $ under various
plausible departures from the benchmark assumptions.

Both our benchmark analysis and the models for our sensitivity analysis
rely on assumptions that identify the conditional distribution of $T$
given the observed data $\mathbf{O}$. The marginal distribution of $T$
is then
obtained as the mixture of the, now identified, conditional
distribution of $%
T$ given $\mathbf{O}$ mixed over the distribution of the observed data
$\mathbf{O}$.

\subsection{Benchmark identifying assumptions}\label{sec3.2}

To help guide our choice of benchmark identifying assumptions, we first note
that since $P[T=k|\mathbf{O}]=0$ if $k\notin{\mathcal T}$ and
$P[T=k|\mathbf{O}]=1$
if $k \in{\mathcal T}$ and $|{\mathcal T}|=1$, we only need
assumptions that suffice to identify $P[T=k|\mathbf{O}]$ when $k \in
{\mathcal T}$ and $|{\mathcal T}|>1$.
Thus, we proceed in reverse order
through the set ${\mathcal T}$ by postulating assumptions that identify
first $P[T=R+1|\mathbf{O}]$ and then sequentially $P[T=k|T \leq
k,\mathbf{O}]$, where
$k \in{\mathcal T}$, $L < k < R+1$. This iterative procedure results
in assumptions that identify $%
P[T=k|\mathbf{O}]$ for all $k\in{\mathcal T}$. This follows because
$P[T=R+1|\mathbf{O}]$ is identified and
for $k \in{\mathcal T}, k < R+1$, $P[T=k|\mathbf{O}]$ equals
\[
P[T \neq R+1|\mathbf{O}] \biggl\{\mathop{\prod_{k < s < R+1}}_{{s \in
{\mathcal T}}} P[T
\neq s|T \leq s,\mathbf{O}] \biggr\} P[T=k|T \leq k,\mathbf{O}].
\]
%

To guide our choice
of benchmark assumptions for identifying $P[ T=r+1|\mathbf{O}=\mathbf
{o}]$, we first note that
given $\mathbf{O}=\mathbf{o}$ the event $T=r+1$ occurs if and only if
the culture at visit~$r$
is positive, that is, if $C_{r}=0$. Our benchmark assumption equates the
unidentified probability $P[ T=r+1|\mathbf{O}=\mathbf{o}] $ with the
identified probability
that $C_{r}=0$ in the subset of patients for which $\mathbf{O}=\mathbf
{o}^{ ( r ) }$,
where $\mathbf{o}^{ ( r ) }$ agrees with $\mathbf{O}$ in all its
components except
that the culture at visit~$r$ is observed. 
Now we write the event $\mathbf{O}=\mathbf{o}$ as the event
\begin{eqnarray}
\label{eq:o}
M_{r}^{c} = 1,
\overline{M}_{r-1}^{c}=\overline{m}_{r-1}^{c},
\overline{C}_{r-1}^{\mathrm{obs}}= \overline{c}_{r-1}^{\mathrm{obs}},
V=v,
M_j^c=0,
C_j=1
\nonumber
\\[-8pt]
\\[-8pt]
\eqntext{\mbox{for } r+1\leq j\leq K,}
\end{eqnarray}
we define the event $\mathbf{O}=\mathbf{o}^{ ( r )}$as the event
\begin{eqnarray}
M_{r}^{c}=0,
\overline{M}_{r-1}^{c}= \overline{m}_{r-1}^{c},
\overline{C}_{r-1}^{\mathrm{obs}}=\overline{c}_{r-1}^{\mathrm{obs}},
V=v,
M_j^c=0,
C_j=1\nonumber\\
\eqntext{\mbox{for }r+1\leq j\leq K,}
\end{eqnarray}
and we postulate that
\begin{equation}
\label{ba1} P[ T=r+1|\mathbf{O}=\mathbf{o}] = P\bigl[ T=r+1|\mathbf{O}=
\mathbf{o}^{ ( r ) }\bigr].
\end{equation}
%

Next we consider assumptions that identify $P[T=k|T\leq k,\mathbf
{O}=\mathbf{o}]$ for $k
\in{\mathcal T}, l < k < r+1$. Once again,
given $ ( T\leq k,\mathbf{O}=\mathbf{o} ) $, the event $T=k$ occurs if
and only if the culture result at visit~$k-1$ is positive, that is, $C_{k-1}=0$.
Because $k \in{\mathcal T}$, $l < k < r+1$, we know that
$C_{k-1}$ is missing, that is, $M_{k-1}^{c}=1$. 
Our benchmark assumption in this case equates
the probability $P[ T=k|T\leq k,\mathbf{O}=\mathbf{o}] $ with the
identified probability that $%
T=k$ (which equates to the event $C_{k-1}=0$) in the subset of
patients for which $\mathbf{O}=\mathbf{o}^{ ( k-1 ) }$, where $\mathbf
{o}^{ ( k-1 ) }$
differs from the subset of patients with $(T \leq k, \mathbf{O}=\mathbf
{o})$ only in that a
sputum culture is observed at visit~$k-1$ and the event $T \leq k$ is
observed to occur (i.e., $M_j^c=0$ and $C_j=1$ for all $k \leq j \leq K$).
Formally, with the event $\mathbf{O}=\mathbf{o}$ defined as in (\ref{eq:o}), we
define the event $\mathbf{O}=\mathbf{o}^{ ( k-1 ) }$ as the event
\begin{eqnarray}
M_{k-1}^{c}=0,
\overline{M}_{k-2}^{c}=\overline{m}_{k-2}^{c},
\overline{C}_{k-2}^{\mathrm{obs}}=\overline{c}_{k-2}^{\mathrm{obs}},
V=v,
M_{j}^{c}=0, C_{j}=1
\nonumber\\
\eqntext{\mbox{for }k\leq j\leq K}
\end{eqnarray}
and assume
\begin{equation}
\label{ba2} P[T=k|T\leq k,\mathbf{O}=\mathbf{o}]=P\bigl[ T=k|\mathbf{O}=
\mathbf{o}^{ ( k-1 ) }\bigr].
\end{equation}
Patients in the subset defined by $ (T\leq k,\mathbf{O}=\mathbf{o} ) $
and subjects in the subset $\mathbf{O}=\mathbf{o}^{ ( k-1 ) }$ have the same
baseline factors and the same recorded history of the auxiliary smear
sputums throughout the study, as well as the same recorded history of
sputum cultures up to visit~$k-2$.

Finally, for a realization $\mathbf{O}=\mathbf{o}$ where $| {\mathcal
T} | > 1$, $P[T=l|T \leq l,\mathbf{O}=\mathbf{o}]=1$.

\subsection{Sensitivity analysis}\label{sec3.3}

The benchmark assumptions (\ref{ba1}) and (\ref{ba2}) are untestable.
For realizations $\mathbf{O}=\mathbf{o}$ with $| {\mathcal T} | >1$, the
following exponential tilt model [\citet{cox1994inference}] expresses
departures from our benchmark
assumptions:
\begin{equation}
\label{sa1} P[T=r+1|\mathbf{O}=\mathbf{o}] = \frac{P[T=r+1|\mathbf
{O}=\mathbf{o}^{ ( r ) }]\exp(\alpha)}{ h_{r+1}(\mathbf{o}^{ ( r )
};\alpha)}
\end{equation}
and for $k \in{\mathcal T}$, $l<k < r+1$,
\begin{equation}
\label{sa2} P[T=k|T\leq k,\mathbf{O}=\mathbf{o}] = \frac{ P[T=k|\mathbf
{O}=\mathbf{o}^{ ( k-1 ) }]\exp(\alpha) }{ h_{k}(\mathbf{o}^{ (k-1 )
};\alpha) },
\end{equation}
where $\alpha$ is fixed and given and $h_{k}(\mathbf{o}^{ ( k-1 )
};\alpha)$ are normalizing constants equal to $E[\exp\{\alpha
I(T=k)\}|\mathbf{O}=\mathbf{o}^{ ( k-1 ) }]$ for $k \in{\mathcal T},
l<k\leq r+1$.
Under this exponential tilt model,
\[
\frac{ \mathrm{odds}(P[T=r+1|\mathbf{O}=\mathbf{o}])}{\mathrm{odds}(P[T=r+1|\mathbf{O}=\mathbf{o}^{ ( r ) }] ) } = \frac{
\mathrm{odds}(P[T=k|T \leq k,\mathbf{O}=\mathbf{o}]) } {\mathrm{odds}(P[T=k|\mathbf{O}=\mathbf{o}^{ ( k-1 ) }] ) }= \exp(\alpha).
\]
Thus, the magnitude of $%
\alpha$ quantifies the departure from our benchmark assumptions.
When $\alpha>0$ ($<0$), $P[T=r+1|\mathbf{O}=\mathbf{o}]$ is greater
(less) than $%
P[T=r+1|\mathbf{O}=\mathbf{o}^{ ( r ) }]$ and $P[T=k|T\leq
k,\mathbf{O}=\mathbf{o}]$ is greater (less) than $P[T=k|\mathbf
{O}=\mathbf{o}^{ ( k-1 ) }]$.
As $\alpha
\rightarrow\infty$ ($-\infty$), $P[T=r+1|\mathbf{O}=\mathbf{o}]$ and
$P[T=k|T\leq
k,\mathbf{O}=\mathbf{o}] $ go to one (zero). When $\alpha\rightarrow
\infty$ ($-\infty$), the ``worst-case'' and ``best-case''
bounds described in the \hyperref[sec1]{Introduction} are attained. $\alpha=0$
corresponds to the benchmark assumption.
To facilitate sensitivity analysis, our class of models assumes that
the departures from the benchmark assumption
are not time-specific.

\subsection{Modeling}\label{sec3.4}

For specified $\alpha$, estimation of the distribution of time of
culture conversion depends on our ability to estimate for each
realization $\mathbf{O}=\mathbf{o}$
with $| {\mathcal T} | >1$, $P[T=k|\mathbf{O}=\mathbf{o}^{ (
k-1 ) }]$ for $k \in{\mathcal T}, l<k\leq r+1$. However, in practice,
these probabilities cannot be estimated
nonparametrically. Therefore, we use a
parametric model for the law of the observed data $\mathbf{O}$ given baseline
cavitation status $X$. This model induces parametric models for
$P[T=k|\mathbf{O}=\mathbf{o}^{ (
k-1 ) }]$, $k \in{\mathcal T}, l<k\leq r+1$, that ultimately enable
estimation of $P[T=k|\mathbf{O}=\mathbf{o}]$ by borrowing information
across strata $%
\mathbf{O}=\mathbf{o}^{ ( k-1 ) }$.

We first recall
that $\mathbf{O}= ( X,\overline{O}_{K} ) $, where $%
O_{k}=(M_{k}^{c},C_{k}^{\mathrm{obs}},M_{k}^{s},S_{k}^{\mathrm{obs}})$ are the data available
at visit~$k$. We model the law of $\mathbf{O}$ by modeling the
distribution of $O_{k}$
given $\overline{O}_{k-1}$ and $X$ for all $k=1,\ldots,K$, where
$\overline{O}_0 = \varnothing$. We use
separate logistic regression models for:
\begin{longlist}[3.]
\item[1.] the probability of $M_{k}^{c}=1$ given $\overline{O}_{k-1}$ and $X$,
that is,%
\begin{equation}
\operatorname{logit}\bigl\{P\bigl[M_{k}^{c}=1|\overline{O}_{k-1},X
\bigr]\bigr\}=a\bigl(k,\overline{O}_{k-1},X;\gamma^{(a)}\bigr),
\label{aa}
\end{equation}

\item[2.] the probability that $C_{k}^{\mathrm{obs}}=1$ given $M_{k}^{c}=0$,
$\overline{O}%
_{k-1}$ and $X$,that is,
\begin{equation}
\operatorname{logit}\bigl\{P\bigl[C^{\mathrm{obs}}_{k}=1|M_{k}^{c}=0,
\overline{O}_{k-1},X\bigr]%
\bigr\}=b\bigl(k,
\overline{O}_{k-1},X;\gamma^{(b)}\bigr), \label{bb}
\end{equation}

\item[3.] the probability that $M_{k}^{s}=1$ given $M_{k}^{c}$,
$C_{k}^{\mathrm{obs}}$, $%
\overline{O}_{k-1}$ and $X$, that is,
\begin{equation}
\qquad\operatorname{logit}\bigl\{P\bigl[M_{k}^{s}=1|M_{k}^{c},C_{k}^{\mathrm{obs}},
\overline{O}%
_{k-1},X\bigr]\bigr\}=c\bigl(k,M_{k}^{c},C_{k}^{\mathrm{obs}},
\overline{O}_{k-1},X;\gamma^{(c)}\bigr), \label{cc}
\end{equation}

\item[4.] the probability that $S_{k}^{\mathrm{obs}}=1$ given $M_{k}^{s}=0$,
$M_{k}^{c}$, $%
C_{k}^{\mathrm{obs}}$, $\overline{O}_{k-1}$ and $X$, that is,
\begin{eqnarray}\label{dd}
&& \operatorname{logit}\bigl\{P\bigl[S^{\mathrm{obs}}_{k}=1|M_{k}^{s}=0,M_{k}^{c},C_{k}^{\mathrm{obs}},
\overline{O}%
_{k-1},X\bigr]\bigr\}
\nonumber\\[-8pt]\\[-8pt]\nonumber
&&\qquad =d\bigl(k,M_{k}^{c},C_{k}^{\mathrm{obs}},
\overline{O}_{k-1},X;\gamma^{(d)}\bigr),
\end{eqnarray}
\end{longlist}
where $a(\cdot)$, $b(\cdot)$, $c(\cdot)$ and $d(\cdot)$ are specified
functions of their arguments and
$\gamma^{(a)}$, $\gamma^{(b)}$, $\gamma^{(c)}$ and $\gamma^{(d)}$ are
unknown parameter vectors.

For the distributions in (\ref{bb}) and (\ref{dd}) we need not consider the
cases $M_{k}^{c}=1$ and $M_{k}^{s}=1$, respectively, because for such settings
the conditional distributions are degenerate ($C_{k}^{\mathrm{obs}}$ is empty
when $%
M_{k}^{c}=1$, and $S_{k}^{\mathrm{obs}}$ is empty when $M_{k}^{s}=1$).

\subsection{Inference}\label{sec3.5}

Under models (\ref{aa})--(\ref{dd}) we can express for all
realizations $\mathbf{O}=\mathbf{o}$
with $| {\mathcal T} | > 1$  the conditional
probability $P[T=k|\mathbf{O}=\mathbf{o}^{ ( k-1 ) }]$ for $k \in
{\mathcal T}$, $l<k\leq r+1$ as given
functions of $\mathbf{o}^{ ( k-1 ) }$ and $\gamma=(\gamma^{(a)},\gamma
^{(b)},\gamma^{(c)},\gamma^{(d)})$ whose expressions, for the special
case where
the right-hand sides of (\ref{aa})--(\ref{dd}) only depend on
$O_{k-1}$, are given in the
\hyperref[appe]{Appendix}. We denote this expression as $P[T=k|\mathbf{O}=\mathbf{o}^{ (
k-1 ) };\gamma]$.
Consequently, if we additionally assume models (\ref{sa1}) and
(\ref{sa2}), then we can express $P[T=r+1|\mathbf{O}=\mathbf{o}]$ and $%
P[T=k|T\leq k,\mathbf{O}=\mathbf{o}]$ ($k \in{\mathcal T}, l < k \leq
r$) as given functions, $P[T=r+1|\mathbf{O}=\mathbf{o};\gamma;\alpha]$
and $%
P[T=k|T\leq k,\mathbf{O}=\mathbf{o};\gamma;\alpha]$ of $\mathbf{o}$,
$\gamma$ and $\alpha$.

The first step is to estimate $\gamma$ using maximum likelihood; denote this
estimator by $\widehat{\gamma}$. This can be done using standard
logistic regression software. This step does not rely on
specification of the sensitivity analysis parameter $\alpha$.

For fixed $\alpha$, we estimate $P[T=k]$ by
$\widehat{P}[T=k;\alpha] = \frac{1}{n} \sum_{i=1}^n \widehat
{P}[T_i=k;\alpha]$,
where $\widehat{P}[T_i=k; \alpha]$ has one of four expressions
depending on $k$ and $\mathbf{O}_i$. If $k \notin{\mathcal T}_i$, then
$\widehat{P}[T_i=k; \alpha]=0$; if $|{\mathcal T}_i|=1$ and $k \in
{\mathcal T}_i$, then $\widehat{P}[T_i=k; \alpha]=1$; if $|{\mathcal
T}_i| > 1$ and $k=R_i+1$, then $\widehat{P}[T_i=k; \alpha] =
P[T=R_i+1|\mathbf{O}=\mathbf{O}_i;\widehat{\gamma};\alpha]$; if
$|{\mathcal T}_i| > 1$ and $k\in{\mathcal T}_i, k<R_i+1$, then
$\widehat{P}[T_i=k; \alpha]$ equals
\begin{eqnarray*}
&& P[T \neq R_i+1|\mathbf{O}=\mathbf{O}_i;\widehat{
\gamma};\alpha] \biggl\{ \mathop{\prod_{k < s < R_i+1}}_{{s \in
{\mathcal T}_i}} P[T \neq s|T\leq
s,\mathbf{O}=\mathbf{O}_i;\widehat{\gamma};\alpha] \biggr\}
\\
&&\qquad{} \times
P[T=k|T \leq k,\mathbf{O}=\mathbf{O}_i;\widehat{\gamma};\alpha].
\end{eqnarray*}

To compare the treatment-specific distributions of time to culture
conversion, one can estimate a common treatment effect over time.
Toward this end, one can use the logistic model for discrete survival
data proposed by \citet{cox1972regression}. This model assumes that
\[
\frac{h_z(k)}{1-h_z(k)} = \tau_k \exp(\beta z),\qquad k=1,\ldots,K, z=0,1,
\]
where $z$ denotes treatment group, $h_z(k) = P_z[T=k|T \geq k]$, and
\mbox{$\tau_1,\ldots,\tau_K \geq0$}. Here $\exp(\beta)$ is the ratio of the
odds of first becoming a culture converter at visit~$k$ given culture
conversion at or after visit~$k$, comparing moxifloxacin with ethambutol.

To estimate the model parameters, one can use equally weighted
minimum-distance estimation [\citet{newey1994large}]. Specifically,
for each choice of $\alpha_{z}$ ($z=0,1$), we minimize the following
objective function:
\[
\sum_{z=0}^1\sum
_{k=1}^K \biggl\{ \frac{\widehat{h}_z(k;\alpha_z) }{ 1- \widehat
{h}_z(k;\alpha_z)} -
\tau_k \exp(\beta z) \biggr\} ^2
\]
with respect to $\tau_{1},\ldots,\tau_{K} \geq0$ and $\beta$, where
$\widehat{h}_z(k;\alpha_z) = \widehat{P}_z[T=k|T \geq k;\alpha_z]$.
For each choice of $\alpha_{z}$ ($z=0,1$), this method finds the
``closest'' fitting logistic model to the ``data:'' $\{\widehat
{h}_z(k;\alpha_z): k=1,\ldots,K, z=0,1 \}$.
Even if the model is incorrectly specified, it can still be used to
provide a valid test of the null hypothesis of no treatment effect.

To estimate the standard error of our estimator, we propose the use of
nonparametric bootstrap.

\section{Data analysis}\label{sec4}

\begin{figure}[t]

\includegraphics{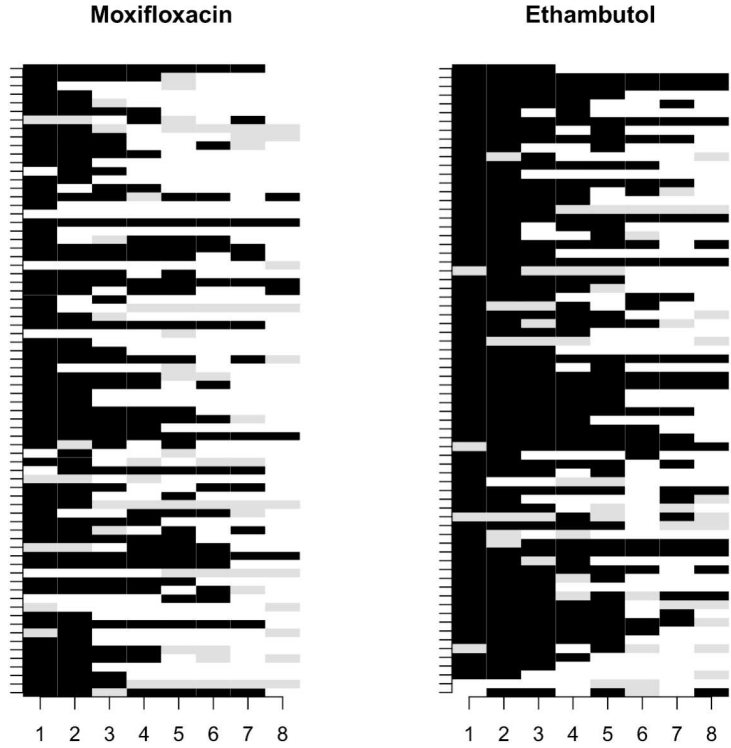}

\caption{Treatment-specific observed culture results, with rows
denoting patients, columns denoting visits, black indicating a positive
culture, white indicating a negative culture and gray indicating a
missing culture.}\label{fig1}
\end{figure}
\begin{figure}

\includegraphics{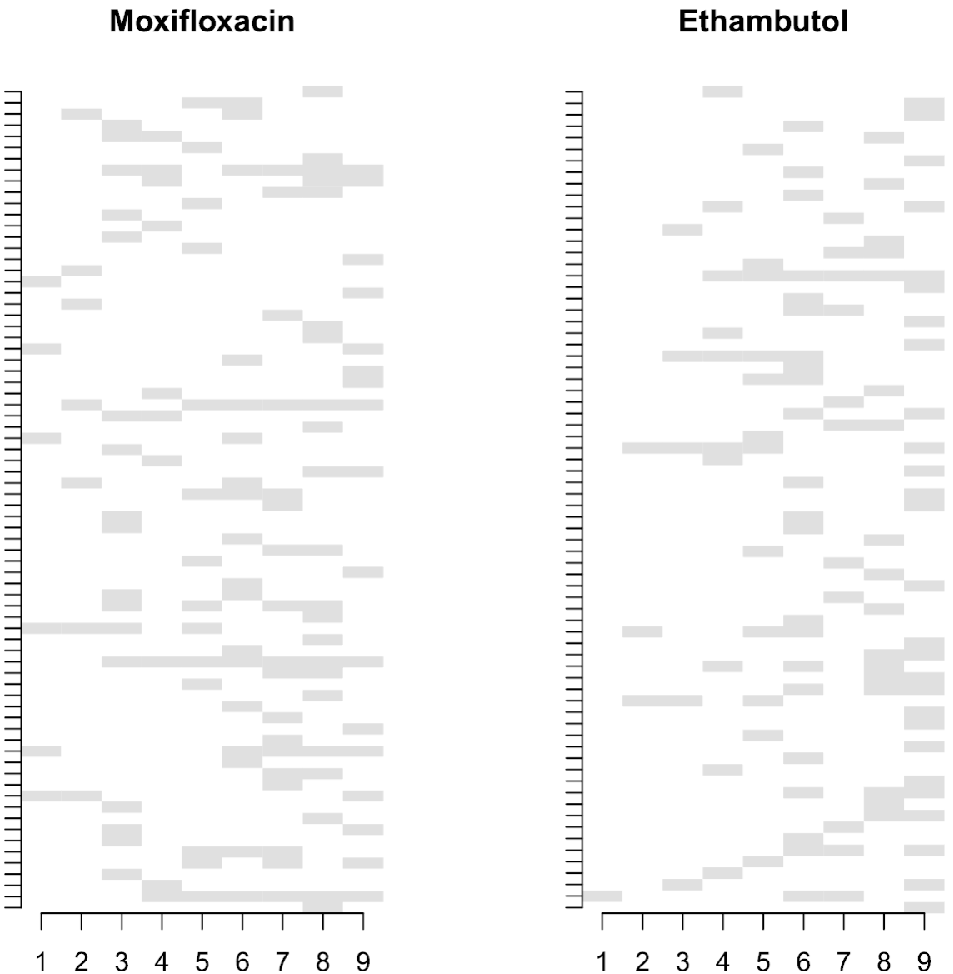}

\caption{Treatment-specific coarsening sets, with rows denoting
patients, columns denoting visits, white indicating infeasible points
and gray indicating feasible points.}
\label{fig2}
\end{figure}

Figure~\ref{fig1} displays the treatment-specific observed culture
results, with rows denoting patients, columns denoting visits, black
indicating a positive culture, white indicating a negative culture and
gray indicating a missing culture. Figure~\ref{fig2} displays the
treatment-specific coarsening sets for time of culture conversion, with
white and gray indicating the infeasible and feasible points, respectively.

%

%

By chance, the treatment groups were not balanced with respect to
cavitation status at baseline;
81.1\% and 56.9\% have cavitation in the moxifloxacin and ethambutol arms,
respectively. It is essential that our analysis adjust for this key
confounder. For each treatment group, we estimate
the distribution of time of culture conversion by a weighted average of
cavitation-specific distribution of time of culture conversion. The weights
are taken to be the marginal (i.e., not conditional on treatment arm)
proportion of patients with and without cavitation at baseline, respectively.

In our data analysis, we considered parsimonious models for the right-hand
sides of (\ref{aa})--(\ref{dd}). Our choice of models was guided by
substantive considerations discussed with our scientist collaborators and
by data analytic model-fitting techniques. Our final model assumed that
the right-hand sides of
\mbox{(\ref{aa})--(\ref{dd})} depended only on
$O_{k-1}=(M_{k-1}^{c},C_{k-1}^{\mathrm{obs}},M_{k-1}^{s},S_{k-1}^{\mathrm{obs}})$ (i.e.,
not on the other components of $\overline{O}_{k-1}$), and the final\vspace*{1pt}
model for
(\ref{cc}) further assumed that
$c(k,M_{k}^{c},C_{k}^{\mathrm{obs}},O_{k-1},X;\gamma^{(c)})
$ did not depend on $C_{k}^{\mathrm{obs}}$ and $C_{k-1}^{\mathrm{obs}}$. The latter
assumption was
imposed because missingness of a sputum culture at visit~$k$ is
highly predictive of missingness of smear sputums at visits $k$ and $k-1$.
In the \hyperref[appe]{Appendix} we show that when the function $%
c(k,M_{k}^{c},C_{k}^{\mathrm{obs}},O_{k-1},X;\gamma^{(c)})$ does not depend on $
C_{k}^{\mathrm{obs}}$ and $C_{k-1}^{\mathrm{obs}}$ for all $k$, $P[T=k|\mathbf{O}=\mathbf
{o}^{ (
k-1 ) } ]$ does not depend on $c(k,M_{k}^{c},C_{k}^{\mathrm{obs}},O_{k-1},X;%
\gamma^{(c)})$ for any $k$, thus alleviating the need to further specify
the function $c(k,M_{k}^{c},C_{k}^{\mathrm{obs}},\break O_{k-1},X;\gamma^{(c)})$.
In the remaining models, we ``borrowed strength'' across treatment groups.
Specifically, we assumed
\begin{eqnarray*}
&& a_z\bigl(k,\overline{O}_{k-1},X;\gamma^{(a)}
\bigr)
\\
&&\qquad  =\gamma_{0,0,k}^{(a)}+\gamma_{0,1,k}^{(a)} X
+ \gamma_{1}^{(a)}I(k>1)M_{k-1}^{c}+
\gamma_{2}^{(a)}I(k>1) \bigl(1-M_{k-1}^{c}
\bigr)C_{k-1}^{\mathrm{obs}}
\\
&&\quad\qquad{} +
\gamma_{3}^{(a)}I(k>1)M_{k-1}^{s}+
\gamma_{4}^{(a)}I(k>1) \bigl(1-M_{k-1}^{s}
\bigr)S_{k-1}^{\mathrm{obs}} + \gamma^{(a)}_5 z,
\\
&& b_z\bigl(k,\overline{O}_{k-1},X;\gamma^{(b)}
\bigr)
\\
&&\qquad = \gamma_{0,k}^{(b)} + \gamma_{1}^{(b)}I(k>1)M_{k-1}^{c}+
\gamma_{2}^{(b)}I(k>1) \bigl(1-M_{k-1}^{c}
\bigr)C_{k-1}^{\mathrm{obs}}
\\
&&\quad\qquad{} +
\gamma_{3}^{(b)}I(k>1)M_{k-1}^{s}+
\gamma_{4}^{(b)}I(k>1) \bigl(1-M_{k-1}^{s}
\bigr)S_{k-1}^{\mathrm{obs}} + \gamma^{(b)}_5 z +
\gamma_6^{(b)} X,
\\
&& d_z\bigl(k,M_{k}^{c},C_{k}^{\mathrm{obs}},
\overline{O}_{k-1},X;\gamma^{(d)}\bigr)
\\
&&\qquad  = \gamma_{0,k}^{(d)} + \gamma_{1}^{(d)}M_{k}^{c}+
\gamma_{2}^{(d)}\bigl(1-M_{k}^{c}
\bigr)C_{k}^{\mathrm{obs}}+\gamma_{3}^{(d)}
\bigl(1-M_{k}^{c}\bigr)C_{k}^{\mathrm{obs}}X
\\
&&\quad\qquad{} +
\gamma_{4}^{(d)}I(k>1)M_{k-1}^{c}+
\gamma_{5}^{(d)}I(k>1) \bigl(1-M_{k-1}^{c}
\bigr)C_{k-1}^{\mathrm{obs}}
\\
&&\quad\qquad{} +\gamma_{6}^{(d)}I(k>1)M_{k-1}^{s}
+
\gamma_{7}^{(d)}I(k>1) \bigl(1-M_{k-1}^{s}
\bigr)S_{k-1}^{\mathrm{obs}} + \gamma_8^{(d)} z +
\gamma_9^{(d)} X,
\end{eqnarray*}
where the functions are subscripted by treatment $z$ ($z=0$ denotes
ethambutol, $z=1$
denotes moxifloxacin).

Tables~\ref{tab2}, \ref{tab3} and \ref{tab4} present estimates of the
exponentiated parameters from these models, along with
95\% nonparametric bootstrap percentile confidence intervals (based on
1000 resamples within treatment groups).
In Table~\ref{tab2}, missingness of sputum culture (aOR${}={}$4.45; 95\% CI:
2.03--9.70) and missingness of smear (aOR${}={}$3.51; 95\% CI: 1.52--9.66) at
a previous visit~are
significant predictors of missingness of sputum culture at the next
visit. In Table~\ref{tab3}, among patients with an observed sputum
culture at visit~$k$, missingness of sputum culture (aOR${}={}$4.00; 95\%
CI: 2.00--9.45), missingness of a smear (aOR${} ={}$5.16; 95\% CI:
1.67--22.47), a negative observed culture (aOR${} ={}$6.37; 95\% CI:
4.08--10.39) and a negative observed smear (aOR${} ={}$3.93; 95\% CI:
2.51--6.18) at visit~$k-1$, as well as assignment to the moxifloxacin
arm (aOR${}= 2.06$; 95\% CI: 1.46--3.19), are significant predictors of
a negative sputum culture at visit~$k$. In Table~\ref{tab4}, among
patients with an observed smear at visit~$k$, missingness of smear (aOR${}={}$3.66; 95\% CI: 1.45--14.53) and an observed negative smear
(aOR${}={}$6.99; 95\% CI: 4.50--11.81) at visit~$k-1$, as well as an observed
negative sputum culture at visit~$k$ (aOR${}={}$10.73; 95\% CI:
5.58--31.18), are significant predictors of a negative smear at visit~$k$. 

%
\begin{table}[t]
\tabcolsep=0pt
\tablewidth=230pt
\caption{(Exponentiated) parameter estimates and 95\% confidence
intervals from the model for missingness of culture results.
See~$a_z(k,\overline{O}_{k-1},X;\gamma^{(a)})$ for form of the model}\label{tab2}
\begin{tabular*}{\tablewidth}{@{\extracolsep{\fill}}@{}lcc@{}}
\hline
\textbf{Intercept} & \textbf{Odds} & \textbf{95\% CI} \\
\hline
Wk 1 $(\exp(\gamma_{0,0,1}^{(a)}))$ & 0.14 & $[0.04,0.30]$\\[2pt]
Wk 2 $(\exp(\gamma_{0,0,2}^{(a)}))$ & 0.11 & $[0.03,0.24]$\\[2pt]
Wk 3 $(\exp(\gamma_{0,0,3}^{(a)}))$ & 0.03 & $[0.00,0.09]$\\[2pt]
Wk 4 $(\exp(\gamma_{0,0,4}^{(a)}))$ & 0.11 & $[0.03,0.25]$\\[2pt]
Wk 5 $(\exp(\gamma_{0,0,5}^{(a)}))$& 0.07 & $[0.01,0.19]$\\[2pt]
Wk 6 $(\exp(\gamma_{0,0,6}^{(a)}))$& 0.04 & $[0.00,0.12]$\\[2pt]
Wk 7 $(\exp(\gamma_{0,0,7}^{(a)}))$& 0.05 & $[0.00,0.14]$\\[2pt]
Wk 8 $(\exp(\gamma_{0,0,8}^{(a)}))$& 0.07 & $[0.01,0.18]$
\\[12pt]
\hline
\textbf{Predictor} & \textbf{Odds ratio} & \textbf{95\% CI}\\
\hline
Wk 1*Cav $(\exp(\gamma_{0,1,1}^{(a)}))$& 0.19 & $[0.00,0.90]$ \\[2pt]
Wk 2*Cav $(\exp(\gamma_{0,1,2}^{(a)}))$& 0.21 & $[0.00,0.87]$ \\[2pt]
Wk 3*Cav $(\exp(\gamma_{0,1,3}^{(a)}))$& 2.93 & $[0.81,\infty]^{\dagger}$\\[2pt]
Wk 4*Cav $(\exp(\gamma_{0,1,4}^{(a)}))$& 0.38 & $[0.09,1.50]$ \\[2pt]
Wk 5*Cav $(\exp(\gamma_{0,1,5}^{(a)}))$& 1.63 & $[0.55,11.28]$\\[2pt]
Wk 6*Cav $(\exp(\gamma_{0,1,6}^{(a)}))$& 1.39 & $[0.36,\infty]^{\dagger}$\\[2pt]
Wk 7*Cav $(\exp(\gamma_{0,1,7}^{(a)}))$& 1.60 & $[0.47,\infty]^{\dagger}$\\[2pt]
Wk 8*Cav $(\exp(\gamma_{0,1,8}^{(a)}))$& 1.63 & $[0.51,8.32]$
\\[3pt]
$I(k >1)M_{k-1}^{c}$ $(\exp(\gamma_{1}^{(a)}))$& 4.45 & $[2.03,9.70]$ \\[2pt]
$I(k>1)(1-M_{k-1}^{c})C_{k-1}^{\mathrm{obs}}$ $(\exp(\gamma_{2}^{(a)}))$& 0.89 & $[0.50,1.68]$\\[2pt]
$I(k>1)M_{k-1}^{s}$ $(\exp(\gamma_{3}^{(a)}))$& 3.51 & $[1.52,9.66]$ \\[2pt]
$I(k>1)(1-M_{k-1}^{s})S_{k-1}^{\mathrm{obs}} $ $(\exp(\gamma_{4}^{(a)}))$& 1.28 & $[0.72,2.28]$ \\[2pt]
Moxifloxacin $(\exp(\gamma_{5}^{(a)}))$& 1.07 & $[0.67,1.76]$ \\
\hline
\end{tabular*}
\tabnotetext[\dagger]{ttt2}{$\infty$ here means a big number.}
\end{table}

\begin{table}[t]
\tabcolsep=0pt
\tablewidth=230pt
\caption{(Exponentiated) parameter estimates and 95\% confidence
intervals from the model for negative culture results. See
$b_z(k,\overline{O}_{k-1},X;\gamma^{(b)})$ for form of the model}\label{tab3}
\begin{tabular*}{\tablewidth}{@{\extracolsep{\fill}}@{}lcc@{}}
\hline
\textbf{Intercept} & \textbf{Odds} & \textbf{95\% CI} \\
\hline
Wk 1 $(\exp(\gamma_{0,1}^{(b)}))$& 0.04 & $[0.02,0.08]$\\[3pt]
Wk 2 $(\exp(\gamma_{0,2}^{(b)}))$& 0.03 & $[0.02,0.06]$ \\[3pt]
Wk 3 $(\exp(\gamma_{0,3}^{(b)}))$& 0.08 & $[0.04,0.14]$ \\[3pt]
Wk 4 $(\exp(\gamma_{0,4}^{(b)}))$& 0.10 & $[0.05,0.18]$\\[3pt]
Wk 5 $(\exp(\gamma_{0,5}^{(b)}))$& 0.10 & $[0.05,0.17]$\\[3pt]
Wk 6 $(\exp(\gamma_{0,6}^{(b)}))$& 0.24 & $[0.13,0.41]$\\[3pt]
Wk 7 $(\exp(\gamma_{0,7}^{(b)}))$& 0.22 & $[0.12,0.40]$\\[3pt]
Wk 8 $(\exp(\gamma_{0,8}^{(b)}))$& 0.49 & $[0.26,0.95]$
\\[12pt]
\hline
\textbf{Predictor} & \textbf{Odds ratio} & \textbf{95\% CI}\\
\hline
$I(k>1)M_{k-1}^{c}$ $(\exp(\gamma_{1}^{(b)}))$& 4.00 & $[2.00,9.45]$ \\[3pt]
$I(k>1)(1-M_{k-1}^{c})C_{k-1}^{\mathrm{obs}}$ $(\exp(\gamma_{2}^{(b)}))$& 6.37 & $[4.08,10.39]$ \\[3pt]
$I(k>1)M_{k-1}^{s}$ $(\exp(\gamma_{3}^{(b)}))$& 5.16 & $[1.67,22.47]$ \\[3pt]
$I(k>1)(1-M_{k-1}^{s})S_{k-1}^{\mathrm{obs}} $ $(\exp(\gamma_{4}^{(b)}))$& 3.93 & $[2.51,6.18]$ \\[3pt]
Moxifloxacin $(\exp(\gamma_{5}^{(b)}))$& 2.06 & $[1.46,3.19]$ \\[3pt]
Cavitation $(\exp(\gamma_{6}^{(b)}))$& 1.16 & $[0.81,1.75]$ \\
\hline
\end{tabular*}
\end{table}

%
\begin{table}[t]
\tabcolsep=0pt
\tablewidth=230pt
\caption{(Exponentiated) parameter estimates and 95\% confidence
intervals from the model for smear results. See
$d_z(k,M_{k}^{c},C_{k}^{\mathrm{obs}},\overline{O}_{k-1},X;\gamma^{(d)})$ for
form of the model}\label{tab4}
\begin{tabular*}{\tablewidth}{@{\extracolsep{\fill}}@{}lcc@{}}
\hline
\textbf{Intercept} & \textbf{Odds} & \textbf{95\% CI} \\
\hline
Wk 1 $(\exp(\gamma_{0,1}^{(d)}))$& 0.25 & $[0.13, 0.42]$ \\[3pt]
Wk 2 $(\exp(\gamma_{0,2}^{(d)}))$& 0.34 & $[0.20,0.56]$ \\[3pt]
Wk 3 $(\exp(\gamma_{0,3}^{(d)}))$& 0.35 & $[0.21,0.58]$\\[3pt]
Wk 4 $(\exp(\gamma_{0,4}^{(d)}))$& 0.21 & $[0.11,0.37]$\\[3pt]
Wk 5 $(\exp(\gamma_{0,5}^{(d)}))$& 0.35 & $[0.19,0.65]$\\[3pt]
Wk 6 $(\exp(\gamma_{0,6}^{(d)}))$& 0.20 & $[0.10,0.39]$\\[3pt]
Wk 7 $(\exp(\gamma_{0,7}^{(d)}))$& 0.23 & $[0.12,0.43]$\\[3pt]
Wk 8 $(\exp(\gamma_{0,8}^{(d)}))$& 0.24 & $[0.11,0.53]$
\\[12pt]
\hline
\textbf{Predictor} & \textbf{Odds ratio} & \textbf{95\% CI}\\
\hline
$M_{k}^{c}$ $(\exp(\gamma_{1}^{(d)}))$& 1.24 & $[0.61, 2.61]$\\[2pt]
$(1-M_{k}^{c})C_{k}^{\mathrm{obs}}$ $(\exp(\gamma_{2}^{(d)}))$ & 10.73 & $[5.58, 31.18]$ \\[2pt]
$(1-M_{k}^{c})C_{k}^{\mathrm{obs}} \cdot\operatorname{Cav}$ $(\exp(\gamma_{3}^{(d)}))$& 0.46 & $[0.16,1.04]$\\[2pt]
$I(k>1)M_{k-1}^{c}$ $(\exp(\gamma_{4}^{(d)}))$& 1.52 & $[0.66,3.46]$\\[2pt]
$I(k>1)(1-M_{k-1}^{c})C_{k-1}^{\mathrm{obs}}$ $(\exp(\gamma_{5}^{(d)}))$ & 1.44 & $[0.92,2.38]$\\[2pt]
$I(k>1)M_{k-1}^{s}$ $(\exp(\gamma_{6}^{(d)}))$ & 3.66 & $[1.45,14.53]$\\[2pt]
$I(k>1)(1-M_{k-1}^{s})S_{k-1}^{\mathrm{obs}} $ $(\exp(\gamma_{7}^{(d)}))$& 6.99 & $[4.50,11.81]$\\[3pt]
Moxifloxacin ($\exp(\gamma_{8}^{(d)})$)& 0.97 & $[0.63,1.48]$\\[2pt]
Cavitation ($\exp(\gamma_{9}^{(d)})$)& 1.18 & $[0.75,1.85]$\\
\hline
\end{tabular*}
\end{table}

\begin{figure}[t]

\includegraphics{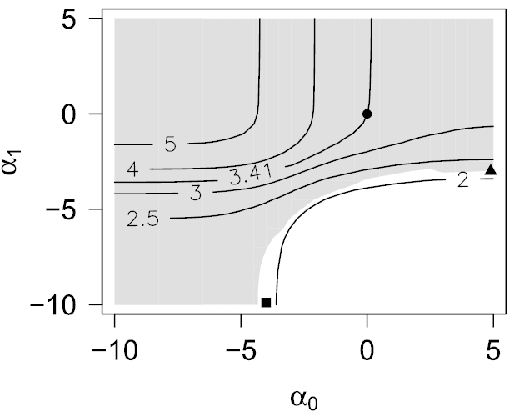}

\caption{Contour plot of the estimated ratio of the odds of first
becoming a culture converter at visit~$k$ given culture conversion at
or after visit~$k$, comparing moxifloxacin vs. ethambutol as a function
of the sensitivity-analysis parameters $\alpha_0$ and $\alpha_1$. The
region in white indicates combinations of $\alpha_0$ and $\alpha_1$
where the lower bound of the 95\% confidence interval is less than 1.
The gray region indicates combinations of $\alpha_0$ and $\alpha_1$
where the null of no treatment difference is rejected in favor of
moxifloxacin. The circle denotes the benchmark assumption $(\alpha
_0=0.0, \alpha_1=0.0)$. The triangle ($\alpha_0=5, \alpha_1=-3$) and
square ($\alpha_0=-4, \alpha_1=-10$) denote two combinations discussed
in the text.}
\label{fig6}
\end{figure}

Under our benchmark assumption, the estimated hazard ratio is 3.41
(95\% CI: $[1.33,13.06]$), indicating that patients treated with
moxifloxacin have a statistically significant shorter time of culture
conversion than those treated with ethambutol. Figure~\ref{fig6}
displays a contour plot of the estimated odds ratio as a function of
$\alpha_0$ and $\alpha_1$. The region in white indicates combinations
of $\alpha_0$ and $\alpha_1$ where the lower bound of the 95\%
confidence interval is less than 1. The gray region indicates
combinations of $\alpha_0$ and $\alpha_1$ where the null of no
treatment difference is rejected in favor of moxifloxacin.

\begin{figure}

\includegraphics{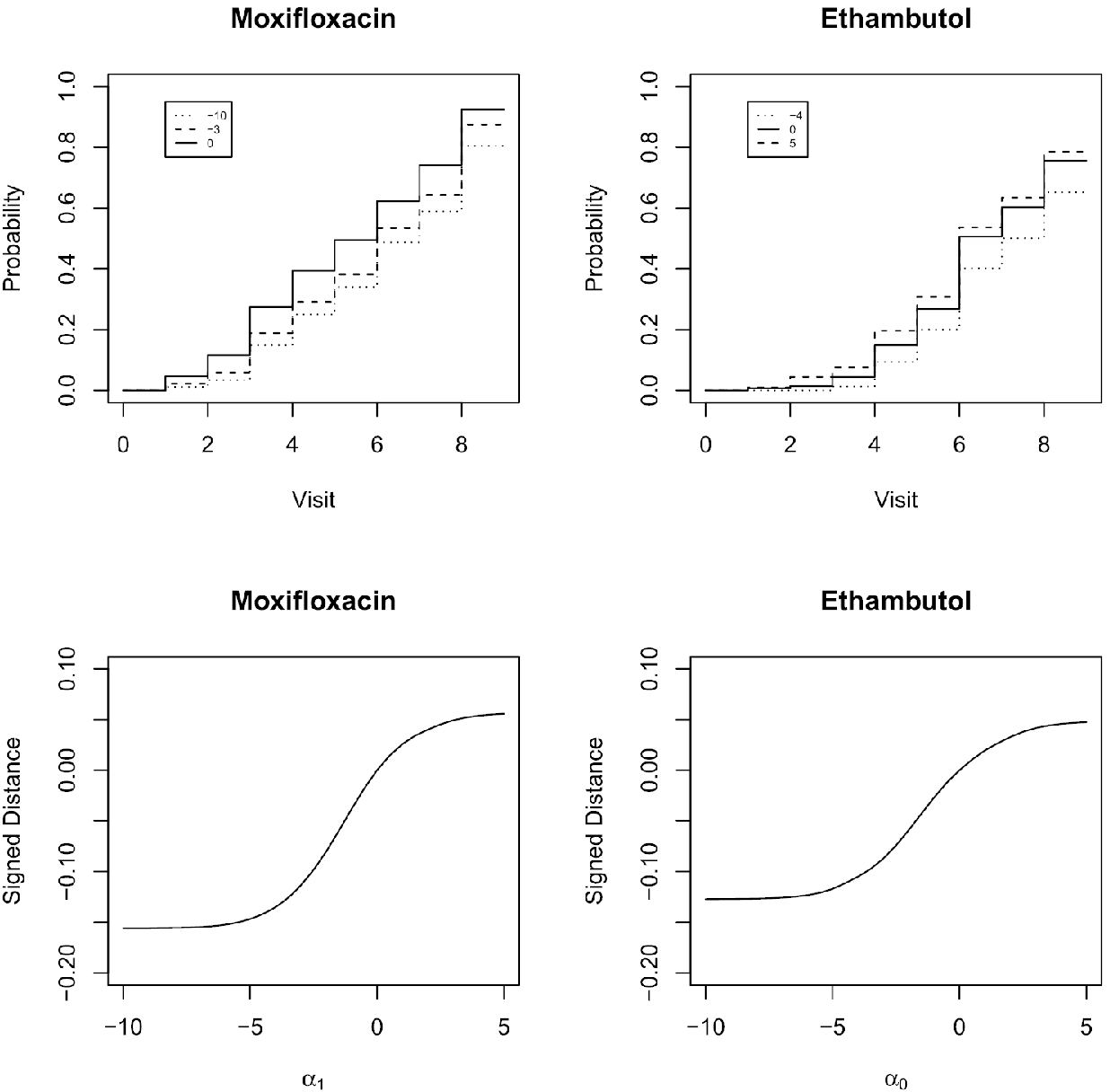}

\caption{First row: Treatment-specific estimated distribution of time
of culture conversion for the benchmark and alternative
sensitivity-analysis parameters considered in the text. Second row: For
each treatment group, the signed Kolmogorov distance between the
estimated distribution of time of culture conversion for given $\alpha$
and the estimated distribution function of time of culture conversion
under the benchmark assumption.}\label{fig5}
\end{figure}

Inference would change relative to the benchmark assumption (circle in
Figure~\ref{fig6}) if, say, $\alpha_0=5.0, \alpha_1=-3.0$ (triangle in
Figure~\ref{fig6}) or $\alpha_0=-4.0, \alpha_1=-10.0$ (square in
Figure~\ref{fig6}).
At these combination of treatment-specific sensitivity-analysis
parameters, the estimated treatment effects are 2.19 (95\% CI:
1.00--14.51) and 2.07 (95\% CI: 0.97--5.38). To understand whether these
combinations are ``far'' from the benchmark assumption, consider
Figure~\ref{fig5}. In the first row, we plot for each treatment group
the estimated distribution of time of culture conversion for these
sensitivity-analysis parameters (dashed and dotted lines) and the
estimated distributions under the benchmark assumption (solid lines).
In the second row, we plot for each treatment group the signed
Kolmogorov distance between the estimated distribution of time of
culture conversion for given $\alpha$ and the estimated distribution
function of time of culture conversion under the benchmark assumption.
The signed Kolmogorov distance for treatment group $z$ with sensitivity
analysis parameter $\alpha_z$ equals
$\widehat{F}_z(k_z^*;\alpha_z) - \widehat{F}_z(k_z^*;0)$, where
\[
k_z^* = \mathop{\operatorname{argmax}}_k \bigl|\widehat{F}_z(k;
\alpha_z) - \widehat{F}_z(k;0) \bigr|
\]
and $\widehat{F}_z(k;\alpha_z)$ is the estimated cumulative
distribution function.
When $\alpha_0=5.0$ and $\alpha_1=-3.0$, the signed distances for the
ethambutol and moxifloxacin arms are $0.047$ and $-0.11$, the latter
being a fairly sizable difference (for the moxifloxacin arm, the
estimated probability of culture conversion by visit~5 is 49.6\% under
the benchmark assumption and 38.2\% when $\alpha_1=-3.0$). Further, the
distances are of opposite signs (i.e., the bias differs between arms).
When we look at other combinations of sensitivity-analysis parameters
where the null hypothesis is not rejected, the sensitivity-analysis
parameter for the moxifloxacin arm is less than or equal to $-3.0$ and
the associated signed distances are at least as extreme as $-0.11$.
When $\alpha_0=-4.0$ and $\alpha_1=-10.0$, the signed distances are
$-0.11$ and $-0.16$ for the ethambutol and moxifloxacin arms,
respectively. Here the signs are in the same direction, but the choice
of sensitivity-analysis parameters yields results that are very close
to the worst-case bounds that assume that all missing cultures are
positive. From a clinical perspective, inferences relative to the
benchmark assumption are fairly robust.


%


\section{Discussion}\label{sec5}

Conde et~al. (\citeyear{conde2009}) did not compare the treatments with respect to time
of culture conversion. Rather, they compared the treatment-specific
probabilities of being a culture converter at or prior to week 8, which
is equivalent to having a negative culture at week 8. They used two
methods. The primary method assumed that all missing cultures at week 8
were positive (moxifloxacin: $77.0\%$; ethambutol: $62.5\%$;
difference: $14.5$ percentage points, 95\% CI [$-0.0$ percentage
points, $29.0$ percentage points]); the secondary method excluded
patients who were missing their week 8 culture, assuming that the
missing cultures were missing completely at random (moxifloxacin:
$90.5\%$; ethambutol: $73.8\%$; difference: $16.7$ percentage points,
95\% CI [$3.4$ percentage points, $30.4$ percentage points]). The
former analysis was not statistically significant, whereas the latter
analysis did suggest a statistically significant treatment effect in
favor of moxifloxacin. Their analysis made no attempt to account for
imbalance in baseline cavitation status between treatment groups.

It is tempting to think that this problem can be addressed by simply
analyzing the culture results using standard statistical methods for
longitudinal binary data (e.g., marginal models and generalized linear
mixed models). A marginal model, fit using generalized estimating
equations, identifies, under the assumption that the culture results
are missing completely at random, the probability of a negative culture
at each visit~$k$; it does not admit identification of the distribution
of time of culture conversion. In contrast, a generalized linear mixed
model is a fully parametric model for the joint distribution of the
culture results and, under the missing at random assumption, admits
identification of the distribution of time of culture conversion.
However, the modeling assumptions are too strong, as they essentially
allow the ``imputation'' of missing culture results that are not needed
to identify the distribution of interest (e.g., the imputation of
missing cultures that are followed by positive cultures). Further, the
model induces testable restrictions, and, as discussed by \citet
{robins1997non}, the missing-at-random assumption is often unrealistic
in follow-up studies with intermittent missing data. Nonetheless, we
fit a logistic-normal generalized linear mixed model to the culture
data with fixed effects for time, treatment and cavitation and a random
intercept. Using the model, we estimated, within levels of time,
treatment and cavitation, the induced probability of a negative culture
among those with missing cultures. Many of the estimated probabilities
were either greater than 1 or less than 0, suggesting inadequate model fit.

An alternative way of analyzing the culture data would treat the data
for each patient as the set of times of culture conversion that are
consistent with their observed culture data (i.e., the coarsening set)
and estimate the distribution of time to culture conversion under the
coarsening-at-random (CAR) assumption [\citet
{gill1997coarsening},
\citeauthor{heitjan1993ignorability} (\citeyear{heitjan1993ignorability,heitjan1994ignorability}),
\citet{heitjan1991ignorability}].
This assumption states that the coarsening process provides no
information about the time of culture conversion beyond conveying that
the true event time is in the observed coarsening set. Under CAR, the
coarsening process is ``ignorable'' (i.e., it factors out of the
likelihood for the observed data). Under this assumption, the estimated
value of $\beta$ is 2.92 (95\% CI: 1.09--11.95). The result is
statistically significant and favors moxifloxacin.

We have assumed, as in most analyses of culture conversion data, that
the test results are measured without error. We know that this is not
correct. It would be interesting to use known information about the
sensitivity and specificity of the culture and smear procedures to
learn about the ``true'' distribution of time of culture conversion.
This will be the subject of future research.

In our analysis, we did not have access to the reasons for missingness.
We know them to be a combination of three main sources: culture
contamination, inability to produce sputum and skipped clinic visits.
For the former two reasons, the culture results are more likely to be
negative. Contamination of sputum cultures with bacteria from the mouth
and airways occurs in 2--10\% of specimens and varies by laboratory.
Patients producing smaller amounts of sputum that is mixed with saliva
are more likely to have contamination; therefore, patients who have
responded to therapy (i.e., have negative cultures) and no longer are
producing large volumes of sputum may be more likely to have a
contaminated specimen. Patients with treated tuberculosis who can no
longer produce sputum are also likely to have responded to therapy and
have negative cultures. If most of the missing data are due to these
two causes, it is not so surprising that the results of the benchmark
analysis are so close to the ``best-case'' bounds.

In summary, we introduced a novel benchmark assumption that allows us
to ``learn'' about the distribution of just those unknown culture
results that are absolutely necessary to identify the distribution of
time of culture conversion by ``borrowing strength'' from patients who
are as similar as possible (with respect to baseline cavitation status,
treatment assignment, and observed culture and sputum results) and on
whom the distribution of these culture results is identified. We
evaluated the sensitivity of inferences to our benchmark assumption by
embedding it in a class of model assumptions indexed by
sensitivity-analysis parameters. Although the sensitivity-analysis
parameters themselves are not scientifically interpretable, the induced
distribution of time of culture conversion (and functionals thereof)
can be estimated and compared with that under the benchmark assumption.
If the differences are judged ``large'' by scientific experts, we hope
that they will comment on the fragility or robustness of the benchmark
inference. Except in rare settings where the treatment effects are so
dramatic or missing data are so minor, we see no alternatives to
sensitivity analysis aided by scientific judgement.

The ideas described in this article can be applied to any study design
in which an enrolled subject is expected to undergo a fixed sequence of
``pass/fail'' tests, one or more test results may be missing, and
interest focuses on estimating the distribution of the earliest test at
which a subject ``passes'' (``fails'') that and all subsequent tests.
For example, the methods described here would be highly relevant for
analyzing studies of treatment of hepatitis C virus (HCV) infection
with antiviral therapy, particularly in light of new and highly active
direct-acting agents. In these studies, patients are typically treated
for 24 or 48 weeks, with HCV viral load measured repeatedly during and
after treatment
[see, e.g., \citet{nelson2012}]. Here a ``pass'' denotes HCV viral
load below the limit of detection. Additionally, the methods can be
easily adapted to address the classic discrete-time interval-censoring
problem where each coarsening set consists of either one time point or
a collection of contiguous time points.

\begin{appendix}
\section*{Appendix}\label{appe}

A straightforward application of the law of total probability entails that
under models (\ref{aa})--(\ref{dd}),
\[
P\bigl[T=k|\mathbf{O}=\mathbf{o}^{ ( k-1 ) };\gamma\bigr]=\frac
{g_{k-1}(0,\overline{O}%
_{k},X;\gamma)}{\sum_{y=0}^{1}g_{k-1}(y,\overline{O}_{k},X;\gamma)},
\]
where
\begin{eqnarray*}
g_{k}(y,\overline{O}_{k+1},X;\gamma)&=& \bigl(1-\operatorname{expit}
\bigl\{a\bigl(k+1,O_{k}(0,y),X;%
\gamma^{(a)}\bigr)
\bigr\}\bigr)
\\
&&{} \times
\operatorname{expit}\bigl\{b\bigl(k,O_{k-1},X;\gamma^{(b)}\bigr)
\bigr\}^{y}
\\
&&{} \times
\bigl(1-\operatorname{expit}\bigl\{b\bigl(k,O_{k-1},X;\gamma^{(b)}
\bigr)\bigr\}\bigr)^{(1-y)}
\\
&&{} \times
\operatorname{expit}\bigl\{b\bigl(k+1,O_{k}(0,y),X;\gamma^{(b)}
\bigr)\bigr\}
\\
&&{} \times
\bigl(1-\operatorname{expit}\bigl\{c\bigl(k,0,y,O_{k-1},X;\gamma
^{(c)}\bigr)\bigr\}\bigr)^{(1-M_{k}^{s})}
\\
&&{} \times
\operatorname{expit}\bigl\{c\bigl(k,0,y,O_{k-1},X;\gamma^{(c)}
\bigr)\bigr\}^{M_{k}^{s}}
\\
&&{}\times
\bigl(1-\operatorname{expit}\bigl\{c\bigl(k+1,0,1,O_{k}(0,y),X;
\gamma^{(c)}\bigr)\bigr\}\bigr)^{(1-M_{k+1}^{s})}
\\
&&{}\times
\operatorname{expit}\bigl\{c\bigl(k+1,0,1,O_{k}(0,y),X;\gamma
^{(c)}\bigr)\bigr\}^{M_{k+1}^{s}}
\\
&&{}\times
\bigl(1-\operatorname{expit}\bigl\{d\bigl(k,0,y,O_{k-1},X;\gamma
^{(d)}\bigr)\bigr\}\bigr)^{(1-S_{k})}
\\
&&{}\times
\operatorname{expit}\bigl\{d\bigl(k,0,y,O_{k-1},X;\gamma^{(d)}
\bigr)\bigr\}^{S_{k}}
\\
&&{}\times
\bigl(1-\operatorname{expit}\bigl\{d\bigl(k+1,0,1,O_{k}(0,y),X;
\gamma^{(d)}\bigr)\bigr\}\bigr)^{(1-S_{k+1})}
\\
&&{}\times
\operatorname{expit}\bigl\{d\bigl(k+1,0,1,O_{k}(0,y),X;\gamma
^{(d)}\bigr)\bigr\}^{S_{k+1}}
\end{eqnarray*}
and $O_{k}(0,y)$ represents observed data at visit~$k$ with $M_{k}^{c}$
set to
0 and $C_{k}$ set to $y$.

If $c(j,M_{j}^{c},C_{j}^{\mathrm{obs}},O_{j-1},X;\gamma^{(c)})$ does not depend
on $C_{j}^{\mathrm{obs}}$ and $C_{j-1}^{\mathrm{obs}}$ for all $j$,
then $P[T=k|O=o^{ ( k-1 ) };\gamma]$ does not depend on $\gamma^{(c)}$.
\end{appendix}

\section*{Acknowledgments}
The authors would like to thank Jonghyeon Kim, Chad Heilig, Malathi
Ram, Pei-Jean Feng and Swarnadip Ghosh for assistance during the
conduct of this research. The authors would also like to thank the
Associate Editor and two anonymous referees  who,  through their detailed
and critical reviews, greatly improved the quality of the manuscript.


%

\printaddresses
\end{document}